\documentclass[sigconf]{aamas}

\usepackage{balance} 

\usepackage{graphicx}
\usepackage{color}
\usepackage{amsmath}
\usepackage{soul}
\usepackage{tcolorbox}
\usepackage{subcaption}
\usepackage{bm}
\usepackage{amsthm}
\usepackage{array}
\usepackage{color}

\usepackage{listings}
\usepackage{xcolor}
\usepackage{tikz}
\usetikzlibrary{shapes,arrows,positioning,calc,fit}
\definecolor{lightblue}{RGB}{173,216,230}
\definecolor{lightgreen}{RGB}{144,238,144}
\definecolor{lightyellow}{RGB}{255,255,224}
\definecolor{mypurple}{RGB}{128, 0, 128}

\setcopyright{ifaamas}
\acmConference[AAMAS '25]{Proc.\@ of the 24th International Conference
on Autonomous Agents and Multiagent Systems (AAMAS 2025)}{May 19 -- 23, 2025}
{Detroit, Michigan, USA}{A.~El~Fallah~Seghrouchni, Y.~Vorobeychik, S.~Das, A.~Nowe (eds.)}
\copyrightyear{2025}
\acmYear{2025}
\acmDOI{}
\acmPrice{}
\acmISBN{}

\acmSubmissionID{119}

\title[AAMAS-2025 Formatting Instructions]{On the limits of agency in agent-based models}

\author{Ayush Chopra}
\affiliation{
  \institution{Massachusetts Institute of Technology}
  \city{Cambridge, MA}
  \country{USA}}

\author{Shashank Kumar}
\affiliation{
  \institution{Massachusetts Institute of Technology}
  \city{Cambridge, MA}
  \country{USA}}

\author{Nurullah Giray Kuru}
\affiliation{
  \institution{Massachusetts Institute of Technology}
  \city{Cambridge, MA}
  \country{USA}}

\author{Ramesh Raskar}
\affiliation{
  \institution{Massachusetts Institute of Technology}
  \city{Cambridge, MA}
  \country{USA}}

\author{Arnau Quera-bofarull}
\affiliation{
  \institution{University of Oxford}
  \city{Oxford}
  \country{UK}}

\begin{abstract}
Agent-based modeling (ABM) offers powerful insights into complex systems, but its practical utility has been limited by computational constraints and simplistic agent behaviors, especially when simulating large populations. Recent advancements in large language models (LLMs) could enhance ABMs with adaptive agents, but their integration into large-scale simulations remains challenging. This work introduces a novel methodology that bridges this gap by efficiently integrating LLMs into ABMs, enabling the simulation of millions of adaptive agents. We present LLM archetypes, a technique that balances behavioral complexity with computational efficiency, allowing for nuanced agent behavior in large-scale simulations. Our analysis explores the crucial trade-off between simulation scale and individual agent expressiveness, comparing different agent architectures ranging from simple heuristic-based agents to fully adaptive LLM-powered agents. We demonstrate the real-world applicability of our approach through a case study of the COVID-19 pandemic, simulating 8.4 million agents representing New York City and capturing the intricate interplay between health behaviors and economic outcomes. Our method significantly enhances ABM capabilities for predictive and counterfactual analyses, addressing limitations of historical data in policy design. By implementing these advances in an open-source framework, we facilitate the adoption of LLM archetypes across diverse ABM applications. Our results show that LLM archetypes can markedly improve the realism and utility of large-scale ABMs while maintaining computational feasibility, opening new avenues for modeling complex societal challenges and informing data-driven policy decisions.\footnote{Corresponding author: ayushc@mit.edu}
\end{abstract}


\newcommand{\BibTeX}{\rm B\kern-.05em{\sc i\kern-.025em b}\kern-.08em\TeX}

\begin{document}


\pagestyle{fancy}
\fancyhead{}


\maketitle 


\section{Introduction}
Many of the today's challenges --- from epidemics to housing shortages to humanitarian crises --- emerge from the complex interplay of countless individuals making decisions and interacting over time. Agent-based models (ABMs) aim to capture these dynamics by simulating collections of agents that act and interact within computational environments. ABMs have proven useful across various domains, including epidemiology \cite{aylett-bullockJuneOpensourceIndividualbased2021b, kerrCovasimAgentbasedModel2021e, hinchOpenABMCovid19AgentbasedModel2021b}, economics \cite{axtellAgentBasedModeling2022, axtell120MillionAgents2016}, and disaster response~\cite{ghaffarian2021agent,aylett2022epidemiological}. For instance, they were used to evaluate vaccination protocols during the COVID-19 pandemic~\cite{romero2021public, aylett-bullockJuneOpensourceIndividualbased2021b,hinchOpenABMCovid19AgentbasedModel2021b}, predict the crash of housing markets~\cite{ge2017endogenous, axtell120MillionAgents2016}, and design evacuation programs for war refugees~\cite{mehrab2024agent,ghaffarian2021agent}. Their utility in addressing policy questions stems from the ability to model interventions by simulating the interplay between individual behaviors and environmental dynamics.

However, the practical application of ABMs has been hindered by two major challenges. First, the high computational costs associated with simulating and calibrating large-scale models have limited their widespread adoption ~\cite{bonabeauAgentbasedModelingMethods2002a}. Recent advancements in deep learning have addressed some of these challenges, enabling the simulation of complex dynamics over millions of agents using vectorized operations 
\cite{chopraDifferentiableAgentbasedEpidemiology2023c, richmondFLAMEGPUFramework2023} and the calibration of models to heterogeneous data sources using differentiable programming~\cite{andelfingerDifferentiableAgentBasedSimulation2023, chopraDifferentiableAgentbasedEpidemiology2023c, montiLearningAgentbasedModels2023b, quera-bofarullBayesianCalibrationDifferentiable2023, tejero-cantero2020sbi, dyerBlackboxBayesianInference2024}. In such differentiable ABMs, deep neural networks also help specify complex environment dynamics~\cite{mordvintsev2020growing}, and autograd facilitates sensitivity analysis in zero-shot \cite{10.5555/3545946.3598853}. Hence, it is now feasible to simulate, calibrate, and analyze ABMs with millions of agents using commodity hardware. Second, and perhaps more critical, is the lack of expressiveness in ABM agents. Many ABMs rely on simplistic, rule-based agent behaviors that fail to capture the nuanced, adaptive decision-making of real-world individuals.

Large language models (LLMs) have shown remarkable performance in text-based applications ~\cite{openaiGPT4TechnicalReport2024,anilPaLMTechnicalReport2023,touvronLLaMAOpenEfficient2023}, suggesting a possible solution to the challenge of creating more realistic agent behaviors. LLM-powered agents have demonstrated potential to enable more general and adaptive human-like behavior~\cite{serapio-garciaPersonalityTraitsLarge2023,hortonLargeLanguageModels2023}. However, integrating LLMs into large-scale ABMs remains problematic. While promising work on multi-agent simulations with LLM-agents has been performed~\cite{parkGenerativeAgentsInteractive2023,vezhnevetsGenerativeAgentbasedModeling2023a}, it has primarily been limited to tabletop games and small population scenarios (few hundred agents). Querying an LLM for each individual agent's decision at every time step quickly becomes computationally infeasible as the number of agents grows into the millions, which is often necessary for studying large-scale complex systems like epidemics or supply chain networks. The goal of our research is to bridge this gap.

\textbf{Contribution:} This work introduces LLM archetypes, a novel methodology that efficiently integrates LLMs into ABMs while maintaining the ability to simulate millions of agents. Our key insight is that by querying LLMs for representative agent types rather than individual agents, we can achieve a balance between behavioral complexity and computational efficiency. LLM archetypes identify representative agent types and use LLM queries to inform the behavior of entire groups of similar agents. Importantly, our approach does not lead to a degenerate solution where all agents within a group make identical decisions. Archetypes maintain action heterogeneity within each group through probabilistic sampling, while significantly reducing the computational burden.





Our approach demonstrates two crucial advantages of LLM archetypes: computational feasibility and enhanced performance. First, we show that LLM archetypes can reproduce population-wide individual behaviors with significantly fewer queries compared to fully LLM-powered agents. This dramatic reduction in computational overhead enables the simulation of millions of agents, which is often necessary for studying large-scale complex systems. Second, LLM archetypes achieve better calibration, and enable flexible counterfactual analysis by preserving simulation scale. We highlight the nuanced trade-off between individual agency and simulation scale, demonstrating that archetypes outperform both fully-adaptive LLM agents (limited to smaller populations due to computational constraints) and simple heuristic agents (which lack adaptivity).

To validate the concept of archetypes, we present a comprehensive analysis using a case study of the COVID-19 pandemic in New York City, simulating 8.4 million agents. This case study showcases how LLM archetypes offer a balanced solution that preserves both adaptive behaviors and computational efficiency. To facilitate utilization across diverse ABMs, we extend AgentTorch - a framework for large-scale agent-based modeling~\cite{chopra2024framework} - to support Archetypes and LLM-powered behaviors.

\section{Background}
In this section, we define the tasks of simulating, calibrating, and analyzing an ABM. We introduce the relevant notation and definitions to formulate the presented experiments. We also formalize the notion of an agent within an ABM and motivate the challenge in scaling LLM-based agent simulations to large populations.

\subsection{Agent-based Modeling}
Consider an ABM composed of $N$ agents. We denote by $\bm s_i(t)$ the state of agent $i$ at simulation time $t$, which contains both static and time-evolving properties of agents. For instance, $\bm s$ may represent the age and sex of a person and their infected status. As the simulation proceeds, an agent $i$ updates their state $\bm s_i(t)$ by interacting with their neighbours $\mathcal N_i(t)$ and their environment $e(t)$, which can both also be time evolving. The neighbourhood of an agent can be specified using a graph, a proximity metric, or other methods. We denote by $m_{ij}(t)$ the message or information that agent $i$ obtains from their interaction with agent $j$. In the case of an epidemiological ABM, this may represent the transmission of infection from agent $j$ to agent $i$. We can then write the agent's update rule as
\begin{equation}
    \label{eq:agent_update}
    \bm s_i(t+1) = f\left(\bm s_i(t), \cup_{j \in \mathcal N(i)} m_{ij}(t), e(t), \bm \theta  \right),
\end{equation}
where $\bm \theta$ are the structural parameters of the ABM. For instance, $\bm \theta$ may correspond to the infectivity of a virus, or the vaccination efficacy. Similarly, the environment can also have its own dynamics that depend on the agent's updates and actions,
\begin{equation}
    \label{eq:env_update}
    \bm e(t+1) = g\left(\bm s(t), e(t), \bm \theta  \right).
\end{equation}
The specific choices of $f$ and $g$ define the dynamics of the ABM system and they are typically stochastic functions which can be mechanistically specified or learned from data. 

\textbf{Simulating an ABM} consists of picking an initial condition for the agents and environment states $(\bm s(0), e(0))$ and recursively applying \autoref{eq:agent_update} and \autoref{eq:env_update}. Despite the very large size of the simulated state space, we are mainly interested in a collection of aggregate outcomes over agent states. 
For most ABMs, this corresponds to a multivariate time-series $\bm x_t = h(\bm s(t))$. For example, in epidemiological ABMs $h$ corresponds to summing over the infected agents to obtain the daily number of infected agents. Once the functional form of an ABM has been set, the simulation of an ABM can be conceived as a stochastic simulator,
\begin{equation}
    \bm x = F(\bm \theta, \bm s(0), \bm e(0)),
\end{equation}
where $F = (f, g) \circ \dots \circ (f, g)$. The composition is repeated for $T$ time-steps. 


\textbf{Calibrating an ABM} refers to the process of finding a set of structural parameters $\hat {\bm \theta}$, or a probability distribution over $\bm \theta$, such that $F(\hat{\bm{\theta}}, \bm s(0), e(0))$ produces an output $\bm{x}$ that is consistent with real-world data $\bm y$. There are various techniques for calibrating ABMs, including approximate Bayesian computation \cite{plattComparisonEconomicAgentbased2020b} and neural likelihood and posterior estimation \cite{dyerBlackboxBayesianInference2024}, among others. 


Once calibrated, we can execute sensitivity \textbf{analyses on ABMs} to understand past events (retrospective), explore alternative scenarios (counterfactual), and design future policies (prospective)~\cite{romero2021public,quera2023don}. This analytical capability makes ABMs powerful tools for policy design and positions them to address the Lucas critique \cite{LUCAS197619}.

\subsection{Scaling Agent-based Models}
Recent advancements, particularly in differentiable ABMs, have made it feasible to simulate, calibrate, and analyze ABMs with millions of agents using commodity hardware. A \textbf{differentiable ABM} ~\cite{chopraDifferentiableAgentbasedEpidemiology2023c, arya2022automatic, quera-bofarullBayesianCalibrationDifferentiable2023} is an ABM for which the gradient
\begin{equation}
\label{eq:diff_abm}
    \eta = \nabla_{\bm\theta} \; \mathbb E[F(\bm \theta)]
\end{equation}
exists and can be computed. This allows ABMs to improve calibration by using gradient-assisted techniques to integrate heterogeneous data\cite{quera-bofarullBayesianCalibrationDifferentiable2023, 10.1145/3604237.3626857}, accelerate simulations on CPUs and GPUs via tensorization~\cite{wscdeepabm}, 
compose with neural networks in end-to-end differentiable pipelines \cite{nca,hypernca,chopraDifferentiableAgentbasedEpidemiology2023c} and accelerate sensitivity analyzes with gradients~\cite{quera2023don}.

AgentTorch~\cite{chopra2024framework} is an open-source framework that allows to generalize these capabilities across diverse ABMs. Its key feature is the ability to differentiate through the simulation ($f, g$), enabling gradient-based optimization of model parameters $\bm \theta$. Through smoothing and reparameterization techniques, AgentTorch achieves differentiability in discrete stochastic programs and allows for the simulation of  tens of millions of agents on consumer-grade GPUs. However, these large-scale simulations have only focused on heuristic. The focus on our work is to preserve this simulation scale while incorporating LLMs to capture adaptive agent behavior. We build on top of AgentTorch for our experiments and analysis. Its flexibility in composing various agent rules and environments makes it a suitable framework for benchmarking agents defined by both heuristic and LLM-based behaviors.

\subsection{Agency in Agent-based Models}
Conventional ABMs typically use heuristic agents update rules $f$ (\autoref{eq:agent_update}) derived from observational data or grounded in theory. However, these rules may not explicitly differentiate between components that depend on agent behaviour and those that depend on the environmental dynamics. An illustrative example of this is the dependence of the probability of infection on the basic reproduction number $R_0$. $R_0$ corresponds to the expected number of cases directly generated by one infected individual. This parameter definition, however, does not allow to distinguish whether a high number of cases is driven by the agent's behaviour (i.e., they interact more), or an increase in the infectivity of the virus (i.e., each contact is more infectious).

Recent advances in Large Language Models (LLMs) have opened new possibilities for creating more realistic and adaptive agent behaviors in ABMs. Integrating LLMs into ABMs can help decouple agent behavior dynamics from environmental dynamics. This modification to the agent update rule (\autoref{eq:agent_update}) is expressed as:
\begin{equation}
    \label{eq:llm_update}
    \bm s_i(t+1) = f\left(\bm s_i(t), \cup_{j \in \mathcal N(i)} m_{ij}(t), e(t), \bm \theta, \ell(\cdot | \bm s_i(t), e(t), \bm \theta)  \right),
\end{equation}
where $ \ell(\cdot | \bm s_i(t), e(t), \bm \theta)$ is the LLM output. For example, when modeling the infection probability of an agent, an LLM could parameterize behaviour related mask-wearing compliance. 
To interpret the LLM output as an action within the ABM environment, we instruct the ABM to return yes / no answers to our prompts. In other words, given an action $\alpha$ (i.e., will the agent isolate at home?) with unknown probability $p$, we use the LLM as a proxy,
\begin{equation}
\label{eq:llm_query}
    \alpha \sim \mathrm{Bernoulli(p)} = \ell (\cdot |   \bm s_i(t), e(t), \bm \theta).
\end{equation}

Several recent works has explored integrating LLMs as ABM agents. Notably, "Smallville"~\cite{parkGenerativeAgentsInteractive2023} simulates 25 LLM-powered agents coordinating to plan a party together, ~\cite{ghaffarzadegan2024generative} simulates disease spread over 100 LLM-powered agents, ~\cite{li2023large} build an expressive environment of macroeconomic dynamics but simulate only 300 agents, ~\cite{yang_project_sid} simulates 1000 LLM-powered agents interacting in minecraft with the aim to capture self-organization in societies.~\cite{vezhnevetsGenerativeAgentbasedModeling2023a} built an ABM framework where both agents and environment are modeled using LLMs. While promising, these works have been limited to small population scenarios (few hundred agents) and not designed to integrate real-world population data~\cite{guo2024large}. Integrating LLMs into large-scale ABMs remains challenging but is necessary for evaluating population-scale complex systems and guiding real-world policy decisions. The focus of our work is to preserve the simulation scale while incorporating LLMs to capture adaptive agent behavior, addressing the critical trade-off between individual agent expressiveness and computational feasibility in large-scale simulations.


\section{LLM archetypes: Scaling LLM as ABM agents}
\label{sec:archetypes}

\begin{figure}
    \centering
    \includegraphics[width=1.0\linewidth]{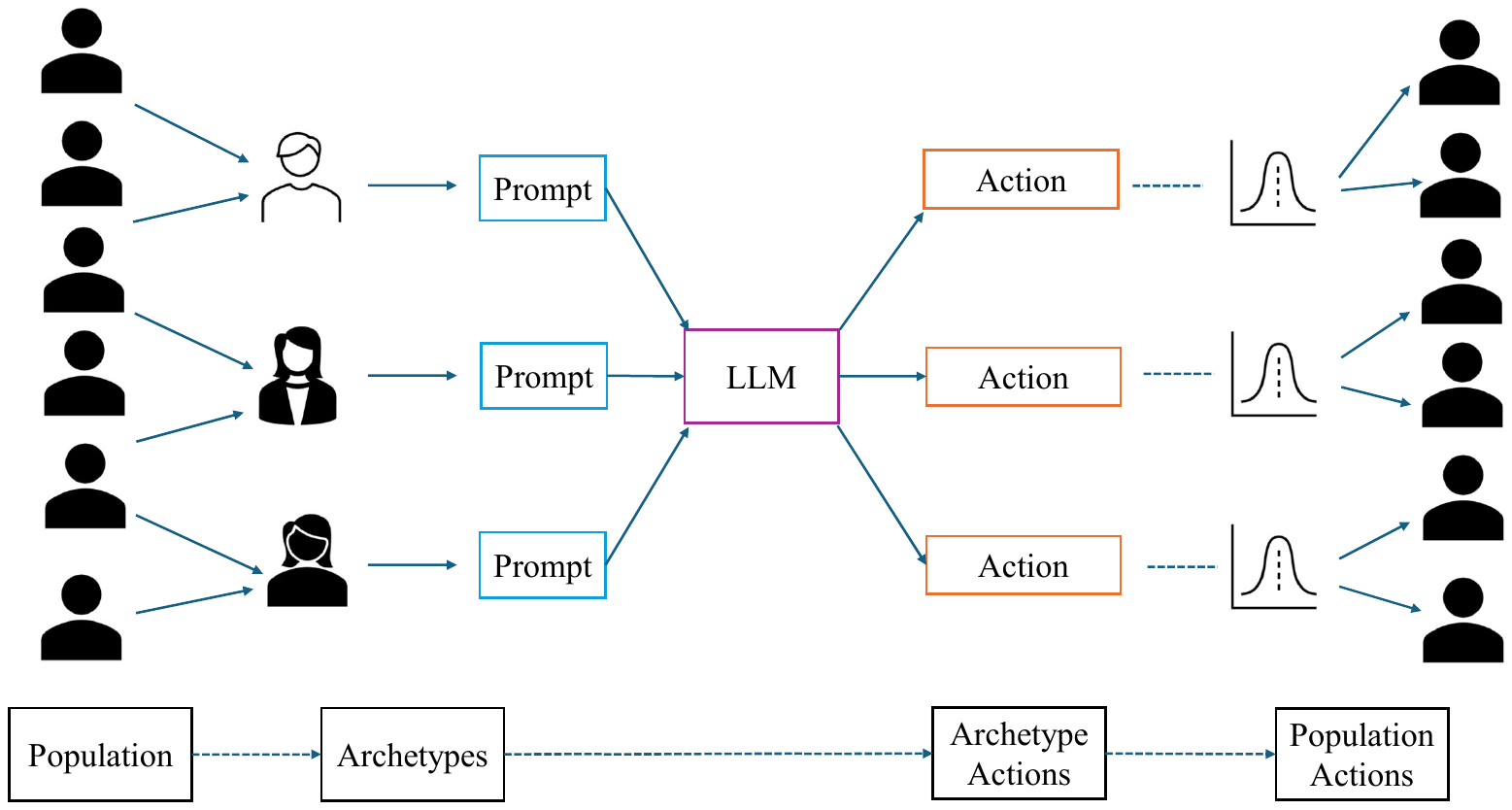}
    \caption{Schematic for sampling individual agent behavior using LLM archetypes. The process involves: (1) assigning individuals to representative archetypes (based on prompt template), (2) querying LLMs for archetype behaviors and estimating action distributions, and (3) sampling individual agent decisions from action distribution of representative archetype. This approach enables efficient scaling of adaptive behaviors to large agent populations.} 
    \label{fig:archetype_viz}
\end{figure}

Understanding complex systems often requires the simulation of the entire population of agents to correctly capture emergent scale-sensitive effects. For instance, while the agency or intelligence of an individual ant may be quite limited, the simulation of the entire colony captures coordination processes wherein ants use themselves as bridges for other ants to use. In these large systems, however, it is infeasible to query \autoref{eq:llm_query} for each agent, time-step, and specific action. This problem can be overcome by recognizing that the number of different behaviors is typically much smaller than the number of agents. In other words, we only need to query the LLM to inform the behaviour over each unique set of agents' characteristics. For instance, if we consider that the behavior is purely informed by age and gender, we only need to consider one LLM query per different combination of age and sex. We refer to each of these unique combinations as \textbf{archetypes}. 

For each possible agent action $\alpha$, we can estimate its probability $p_\alpha$ in \autoref{eq:llm_query} using Monte-Carlo,
\begin{equation}
\label{eq:p_arch}
\begin{split}
    p_\alpha(\bm {s_i}(t), e(t), \bm \theta) &= \mathbb E \left [\ell (\cdot |   \bm s_i(t), e(t), \bm \theta) \right]\\
    &\approx \frac{1}{M} \sum_{j=1}^M \xi_i \;\text{ with } \; \xi_i\sim \ell(\cdot |  s_i(t), e(t), \bm\theta).
\end{split}
\end{equation}
By estimating $p_\alpha(k)$ for each archetype $k$, we can simulate the action of its agent by sampling the action from the archetype to which it belongs. Let $K$ be the number of agent archetypes and $A$ be the number of LLM-queryable actions; we can then simulate the behaviour of all agents with $K\times A$ queries, which will be typically much smaller than the number of agents $N$, allowing us to scale the simulation to millions of agents. This is shown in ~\autoref{fig:archetype_viz}.

\section{Experimental Setup}
\label{sec:motivation}
This section formalizes the simulation environment we use to evaluate LLM archetypes and analyze the trade-off between agency and simulation scale. We consider a large-scale ABM of New York City during the COVID-19 pandemic, simulating 8.4 million individuals in a complex real-world scenario.

The COVID-19 pandemic exemplifies the intricate interplay between individual behavior, policy interventions, and environmental factors that our approach aims to capture. Disease spread initially triggered fluctuating mobility patterns and multiple infection waves~\cite{santana2023covid}. Government lockdowns, while controlling spread, caused severe economic consequences, including unprecedented unemployment~\cite{dreger2021lockdowns}. Stimulus programs, introduced to mitigate economic hardship and encourage compliance to health measure~\cite{li2021impacts}, had unintended effects on labor markets and resource allocation~\cite{falcettoni2020acts}. As the pandemic progressed, "pandemic fatigue" emerged, further complicating public health compliance and economic recovery~\cite{qin2023pandemic}. This complex feedback loop between health outcomes, economic conditions, and human behavior provides an ideal environment to demonstrate the benefit of capturing and simulating adaptive agent behaviors at scale.

\textbf{Environment:} The agent states have static (age, gender, income, occupation) and dynamic (disease, employment status) attributes. We use 2022 American community survey (ACS) for demographic and household characteristics, the Bureau of Labor Statistics for employment data, and the center for disease control (CDC) reports for data on disease dynamics, consistent with prior work~\cite{romero2021public, wscdeepabm, hinchOpenABMCovid19AgentbasedModel2021b}. Agent attributes are specified at census-resolution with demographic and income information discretized into bins. 
Interactions occur over household, workplace, and mobility networks, with recreational and workplace mobility parameterized using Google Mobility trends. Our simulations focus on the dynamics of disease spread and labor market. For disease spread, we consider a standard epidemiological model~\cite{hinchOpenABMCovid19AgentbasedModel2021b, chopraDifferentiableAgentbasedEpidemiology2023c, aylett-bullockJuneOpensourceIndividualbased2021b} wherein infection spreads through contact and the probability of agent $i$ getting infected at step $t$ is:
\begin{equation}
    \label{eq:infection_base}
    p_i(t) = 1 - \exp\left(-\frac{\beta\, S_i}{n_i} \sum_{j \in \mathcal N(i)} I_j(t)\right),
\end{equation}
where $\mathcal N(i)$ is the set of neighbors of agent $i$, $S_i$ the susceptibility of agent $i$, $I_j$ the infection status of each neighbour, $n_i = \# \mathcal N(i)$ the total number of neighbors, and $\beta$ a structural parameter of the ABM called the effective contact rate. The neighbourhood $\mathcal N(i)$ is given by a contact network constructed from household and mobility data in the US census.

For labor market, we consider a standard econometric model~\cite{li2023large} which relates participation behavior of individual agents with aggregate unemployment rate at time $t$ ($\mu_{w,t}$) as:
\begin{equation}
\label{eq:work_base}
    \mu_{w,t} = \sum_{j \in \mathcal{N}} \gamma_{0} W_{j}(t) + \gamma_{1}C_{t}
\end{equation}
where $W_j(t)$ is the willingness to work for agent $j$ at time $t$ and $C_{t}$ is the history of unemployment claim rates in the region, obtained from census data; and $\gamma_{0}$ and $\gamma_{1}$ are the structural parameters.

The epidemiological model forecast cases, while the economic model forecast unemployment rates. The models are coupled through a feedback loop: case numbers affect agents' willingness to work, which influences labor-force participation rates and workplace interaction networks, which in turn affect disease transmission. We implement this environment using the AgentTorch framework~\cite{chopra2024framework} which enables differentiate through these stochastic dynamics and scales to large populations (8.4 million agents). The parameters ($\beta, \gamma_0, \gamma_1$) are calibrated to real-world data for cases and unemployment rates, using a standard protocol for differentiable ABMs~\cite{chopraDifferentiableAgentbasedEpidemiology2023c,quera-bofarullBayesianCalibrationDifferentiable2023,chopra2024framework} (visualized in figure~\ref{fig:calibration_tikz}). More details are in appendix A.

\textbf{Behavior}: We use LLMs to model isolation and employment behavior of individual agents. Our prompt includes agent demographics, disease dynamics, information about extrinsic interventions (stimulus payment) and intrinsic behavior adaptation (duration of pandemic to capture effect of "pandemic fatigue"). The user prompt template, motivated by~\cite{li2023large}, is given below:

\begin{tcolorbox}[colback=green!5!white, colframe=blue!55!black, title=User Prompt]
You are a \{gender\} of age \{age\}, living in the \{location\} region and receiving a monthly income of \{income\}. 
\par\vspace{0.1cm}
The number of new cases in your neighborhood is \{cases\}, which is a \{change\}\% change from the previous month. It has been \{duration\} months since the start of the pandemic.
\par\vspace{0.1cm}
This month, you have received a stimulus payment of \{payment\} to support your living expenses.

\par\vspace{0.1cm}
Given these factors, do you choose to isolate at home? (isolation behavior)

\par\vspace{0.1cm}
Given these factors, do you choose to work? (employment behavior)

\par\vspace{0.1cm}
"There isn't enough information" and "It is unclear" are not acceptable answers.
Give a "Yes" or "No" answer, followed by a period. Give one sentence explaining your choice.
\end{tcolorbox}

The input prompt receives case numbers and pandemic duration from the past simulation trajectory, instead of ground-truth data, and outputs agents' willingness to work $W_j(t)$ which is further used in simulation. We conduct such auto-regressive prompting for two reasons: i) when simulating for prospective interventions, ground-truth data is not available and hence prompt needs to be specified entirely using simulation, ii) when simulation is un-calibrated, the model peaks may not align well with real-world data. In such case, using ground-truth data is unsuitable for capturing the adaptability of behavior (especially when incorporating time-varying information like infections). Since prompt at step $t$ depends upon simulated trajectory at step $t-1$, LLMs need to be sampled online during the simulation. As behavior cannot be sampled offline, the trade-off between simulation scale and agent behavior becomes particularly critical. Finally, the stimulus eligibility, timing and amounts are based on the policy implemented in NYC at the time. Specifically, December 2020 - March 2021 overlaps with the second stimulus check which provided adults \$ 600 and additional \$ 600 for every child~\cite{nyc311}. We use GPT-3.5 for our experiments. The system prompt provides context about the disease dynamics, relative susceptibility of different demographics and is provided in Appendix A. 

\begin{figure*}[t!]
    \centering
    \includegraphics[width=0.95\linewidth]{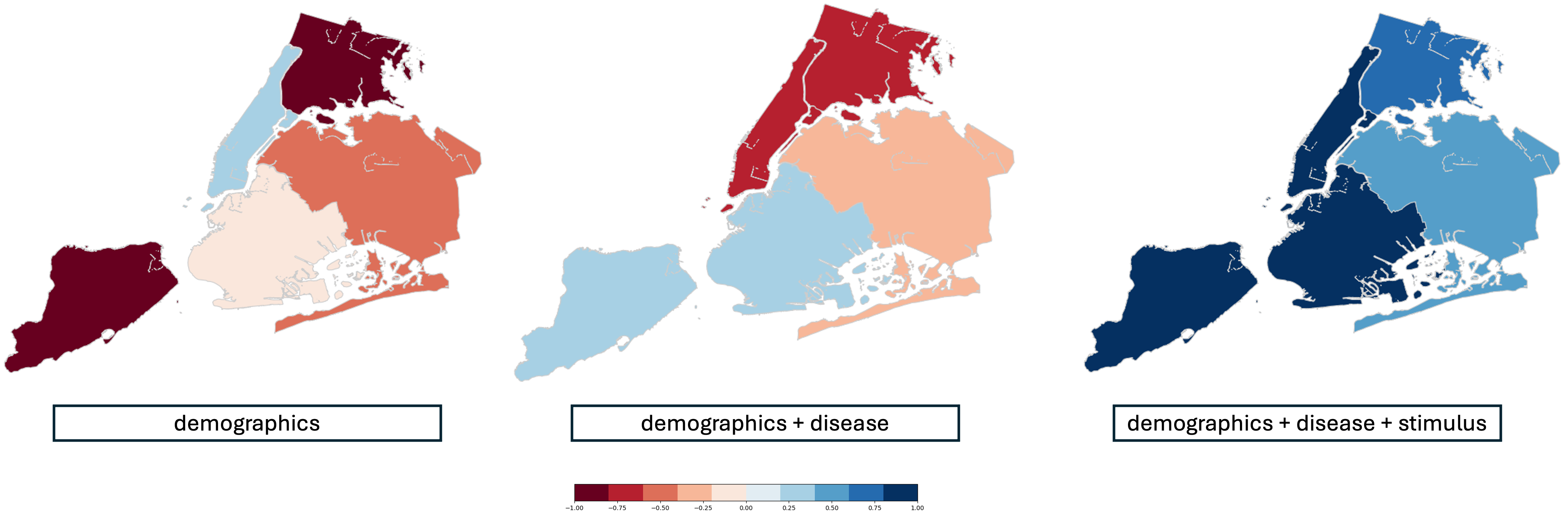}
    \caption{\textbf{Prompting agents via LLM archetypes}: Correlation between population-wide employment behavior predicted by LLM archetypes and observed data for 8.4 million NYC agents. Prompt 1 (left) corresponds to scenario where LLMs only see demographic attributes. Prompt 2 (middle) and Prompt 3 (right) add further contextual information regarding disease dynamics and stimulus payments. Increased correlation with additional contextual information highlights the ability of LLMs to capture behaviour trends across demographics and geography.}
    \label{fig:geoplots-space}
\end{figure*}


\section{Validating Archetypes}
\label{sec:benchmark_llm_agents}
The section benchmarks the capacity of Archetypes to prompt behavior consistent with measurable population-wide observations. We use LLM archetypes to prompt individual-level willingness to work for 8.4 million synthetic agents (consider responses only for eligible working adults), focusing on two time-periods: December 2020 to March 2021 (coinciding with the second stimulus round) and March to May 2022 (post-Omicron wave). We test three scenarios with increasing contextual information: (Prompt 1) we only provide demographic attributes of the agent, (Prompt 2) we add information about disease dynamics, (Prompt 3) we further include information about access to stimulus payments. For each scenarios, we initialize 3 ($M$ in equation~\ref{eq:p_arch}) queries per archetypes representing different combinations of unique prompts over the considered time frame. Given the census-resolution (demographic and income attributes binned for privacy) and prompt design (disease dynamics and stimulus information shared by all agents), this approach requires only \~400 LLM queries to sample one decision (weekly) for each individual in the population. This is a significant reduction compared to millions of queries required in the conventional paradigm.

Following \autoref{eq:p_arch}, we obtain the probability of each archetype $k$ performing action $\alpha$ ("will you work?") for each week, $p_\alpha(k, t)$. By sampling from the induced Bernoulli distributions, we generate a time-series of work attendance. We aggregate these individual-level actions to calculate the change in labor force participation rate across New York boroughs and compute correlation of this time series with observed data from the US Bureau of Labor Statistics \footnote{https://www.bls.gov/charts/employment-situation/civilian-labor-force-participation-rate.htm}. Results are averaged across 5 independent runs for robustness.

\autoref{fig:geoplots-space} shows increasing correlation between behavior generated by LLM-archetypes and census data as we add more contextual information to the prompt. This demonstrates the method's ability to incorporate nuanced factors - like evolving environments and incentives - into agent decision making. Notably, archetypes capture positive time-varying correlations in census-level behaviors for 3 of 5 boroughs (roughly 5 million people across income and demographic), which is an encouraging result. 

In the second experiment (March to May 2022), we test the ability of LLM-archetypes to simulate adaptive behavior over time, particularly "pandemic fatigue". This period represents the scenario post-Omicron wave, when stimulus and unemployment payments had also declined considerably. We modify the prompt to highlight the time duration since the start of the pandemic and the lack of financial incentives, testing whether our representative (and individual) agents are sensitive to these changes. We then repeat the population sampling procedure described in the previous experiment to obtain the time-series and examine cross-correlation with real-world data on participation rates. Results are presented in Appendix B and highlight the ability of LLM-archetypes to capture how individual behavior adapts over time, which is encouraging for use in agent-based modeling.

This analysis demonstrates the potential of LLM archetypes to capture complex, adaptive behaviors in large-scale populations while maintaining computational efficiency. Despite high-variance individual responses, archetypes successfully capture positive time-varying correlations with census-level behaviors, an encouraging outcome for large-scale ABMs. To mitigate biased LLM responses, we estimation archetype distributions via multiple LLM generations ($M$ queries per archetype), as motivated by~\cite{manakul2023selfcheckgpt}. While these results are promising, we acknowledge ongoing challenges. First, real-world behaviors can be significantly more complex than what our prompt can capture and second, data contamination in LLMs remains an open challenge with no formal technique to design prompts for LLM queries. Despite these limitations, the computational efficiency of archetypes - requiring only 400 queries for 8.4 million agents - compared to millions in the conventional approach represents a significant advance. This efficiency, combined with the ability to capture adaptive behaviors, makes LLM archetypes a promising tool for large-scale ABMs. We demonstrate the viability of this approach for large-scale simulations in Section~\ref{sec:calibration}, presenting performance benchmarking results of online LLM sampling coupled with a highly complex environment.
\section{Behavior Agency vs Simulation Scale}
\label{sec:calibration}
This section investigates the impact of incorporating adaptive agent behavior within ABMs and analyzes the trade-off between agency and simulation scale when calibrating to real-world data. We simulate dynamics of disease spread and labor market in New York City from December 2020 to April 2021 and, compare three kinds of agent architectures: LLM-as-agent, Heuristics and Archetypes. LLM-as-agent instantiates a unique LLM query per individual agent, Archetype instantiates $M$ LLM queries per representative agent and heuristic agents used hand-crafted behaviors shared across all agents. Specifically, heuristic agents use \autoref{eq:infection_base} and \autoref{eq:work_base} as they are. For archetype and LLM-as-agent, these equations are modified to incorporate an action $\alpha$ determined by the LLM output and obtain agent decision to isolate ($I_{j}$) and work ($W_{j}$). In terms of $I_{j}$, this is defined as:
\begin{equation}
    \label{eq:infection_llm}
    I_j^\mathrm{LLM} := I_j(t) (1-\ell_\alpha(\bm s_j(t), e(t))),
\end{equation}
where $\ell_\alpha$ is the LLM output for the action, and
\begin{equation}
    \label{eq:infection_arch}
    I_j^\mathrm{archetype}(t) := I_j(t) (1-A_j(t)),
\end{equation}
where $A_j(t)$ is sampled from $\mathrm{Bernoulli}(p_j(\alpha))$ with $p_j(\alpha)$ is estimated using the LLM (see \autoref{sec:archetypes}). Similarly, $W_j(t)$ is also modified for LLM-as-agent and archetypes.

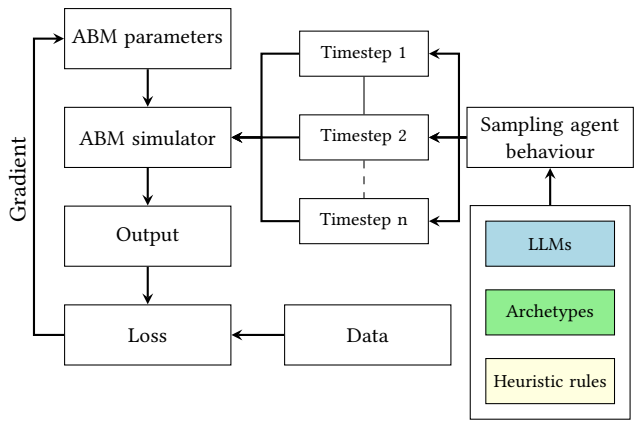
\begin{figure}[h!]
\begin{tikzpicture}[
    box/.style={rectangle,draw,text width=2cm,text centered,minimum height=0.8cm,font=\small},
    colorbox/.style={rectangle,draw,text width=1.5cm,text centered,minimum height=0.6cm,font=\footnotesize},
    timebox/.style={rectangle,draw,text width=1.5cm,text centered,minimum height=0.6cm,font=\footnotesize},
    arrow/.style={->,>=stealth,thick},
    node distance=0.5cm and 0.7cm
    ]
\node[box] (params) {ABM parameters};
\node[box,below=of params] (simulator) {ABM simulator};
\node[box,below=of simulator] (output) {Output};
\node[timebox,right=0.9cm of simulator] (timestep2) {Timestep 2};
\node[timebox,above=0.5cm of timestep2] (timestep1) {Timestep 1};
\node[timebox,below=0.5cm of timestep2] (timestepn) {Timestep n};
\node[box,right=0.5cm of timestep2] (sampling) {Sampling agent behaviour};
\node[colorbox,fill=lightblue,below=0.7cm of sampling] (llms) {LLMs};
\node[colorbox,fill=lightgreen,below=0.3cm of llms] (archetypes) {Archetypes};
\node[colorbox,fill=lightyellow,below=0.3cm of archetypes] (rules) {Heuristic rules};
\node[box,below =of output] (loss) {Loss};
\node[box,right=of loss] (data) {Data};
\draw[arrow] (params) -- (simulator);
\draw[arrow] (simulator) -- (output);
\draw[arrow] (output) -- (loss);
\draw[arrow] (data) -- (loss);
\draw[arrow] (sampling) -- (timestep2);
\draw (timestep1) -- (timestep2);
\draw[dashed] (timestep2) -- (timestepn);
\draw[arrow] (timestep1.west) -| ++(-0.5,0) |- (simulator.east);
\draw[arrow] (timestep2.west) -| ++(-0.5,0) |- (simulator.east);
\draw[arrow] (timestepn.west) -| ++(-0.5,0) |- (simulator.east);
\draw[arrow] (sampling.west) -- ++(-0.1,0) |- (timestep1.east);
\draw[arrow] (sampling.west) -- ++(-0.1,0) |- (timestep2.east);
\draw[arrow] (sampling.west) -- ++(-0.1,0) |- (timestepn.east);
\node[draw,fit=(llms) (archetypes) (rules),inner sep=0.2cm] (container) {};
\draw[arrow] (container.north) -- (sampling.south);
\draw[arrow] (loss.west) -| ($(loss.west)-(.4,0)$) |- node[pos=0.35, left=0.22cm, rotate=90] {Gradient} (params.west);
\end{tikzpicture}
\caption{Calibration protocol for the three types of agent behaviours considered. This involves simulating ABM by sampling agent behavior at each step, comparing outputs to real-world data, and adjusting parameters through gradient-based optimization. More details are in Appendix C.}
\label{fig:calibration_tikz}
\end{figure}

When simulating, we sample agent decisions at each step to execute dynamics and, aggregate infection and employment states after $N$ steps. The aggregated outputs are used to calibrate structural parameters $\bm \theta = (\beta, \gamma_0, \gamma_1)$ to historical time-series of infections and unemployment rates, using the protocol visualized in ~\autoref{fig:calibration_tikz}. LLMs-as-agents query behavior at individual level and hence using it to simulate 8.4 million agents is computational infeasible. This enforces a trade-off between simulation scale and individual agency. For fair comparison, we fix the prompt budget to 300 LLM queries per step and compare the following configuration: a) Heuristic agents simulate 8.4 million agents using hand-crafted behaviors, b) Archetypes simulate 8.4 million agents with representative-level LLM queries and c) LLM-as-agents simulate a smaller population of 300 agents with individual-level LLM queries. Output of LLM-as-agents are scaled to the full population to compare with historical data and evaluate performance. To evaluate each calibrated model, we simulate a future time-series of 16 weeks for infection data (measured weekly) and 80 weeks for employment rates (measured monthly) and report forecasting errors.


\begin{figure}[h!]
    \centering
    \includegraphics[width=0.85\linewidth]{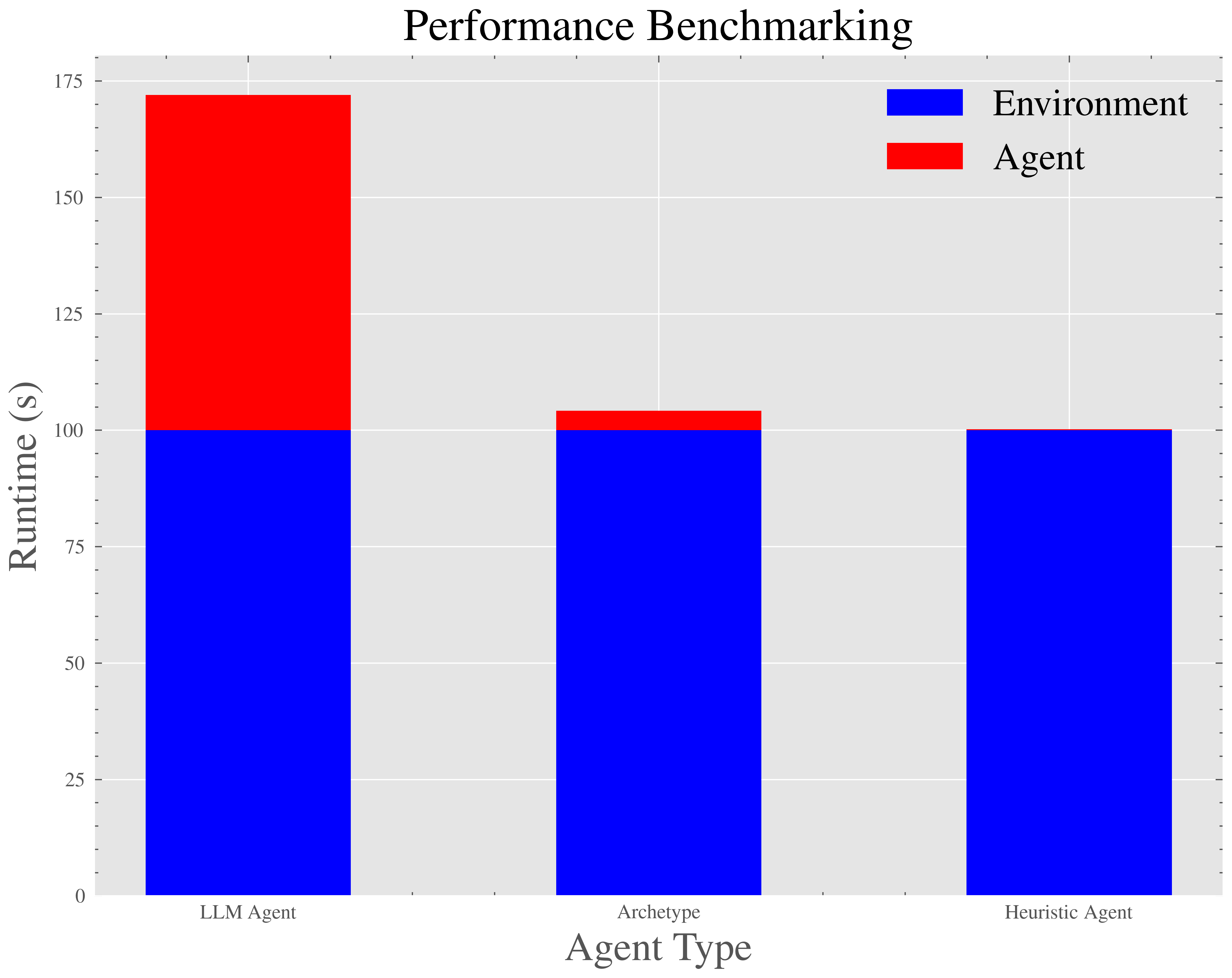}
    \caption{Runtime benchmarks for the environment and agent. Archetypes introduce much lower runtime overhead, enabling the simulation to scale to larger population size}
    \label{fig:runtime-performance}
\end{figure}

\textbf{Analysis}: Results presented in ~\autoref{tab:error_rates} show that archetype-based model achieves the best performance, highlighting both the need for adaptive and expressive agents and the requirement of simulating the entire scale of the system. Further, heuristic agents simulated at population scale outperform LLM agents constrained to small population samples which shows that computational scaling is crucial. The benefit of LLM-archetypes over heuristic agents shows the benefit of capturing behavioral adaptations can be extremely useful at the right simulation scale. ~\autoref{fig:runtime-performance} shows that Archetypes achieve this superior performance while consuming 95\% less run-time compared to LLM-as-agents and marginally more than heuristic agents, which is encouraging for practical utility.



\begin{table}[h!]
\begin{tabular}{|c|c|c|}
\hline
Agent | Error Rate   & Unemployment  ($\downarrow$) & Infection ($\downarrow$)                   \\ \hline
Archetype    & \textbf{24.59 $\pm$ 1.5} & \textbf{95.17 $\pm$ 20.23}    \\ \hline
Heuristic    & 41.05 $\pm$ 0.1 & 2914.73 $\pm$ 300.25 \\ \hline
LLM-as-agent & 56.98 $\pm$ 2.5 & 4311.70 $\pm$ 674.14 \\ \hline
\end{tabular}
\caption{Benchmark results showing the mean-square errors for each of the considered agent architectures. Archetype achieves lower test error when forecasting both infections and unemployment rates as they capture adaptive agent behaviors \textit{without} compromising simulation scale.}
\label{tab:error_rates}
\end{table}

\begin{figure*}[t!]
    \centering
    \includegraphics[width=0.95\linewidth]{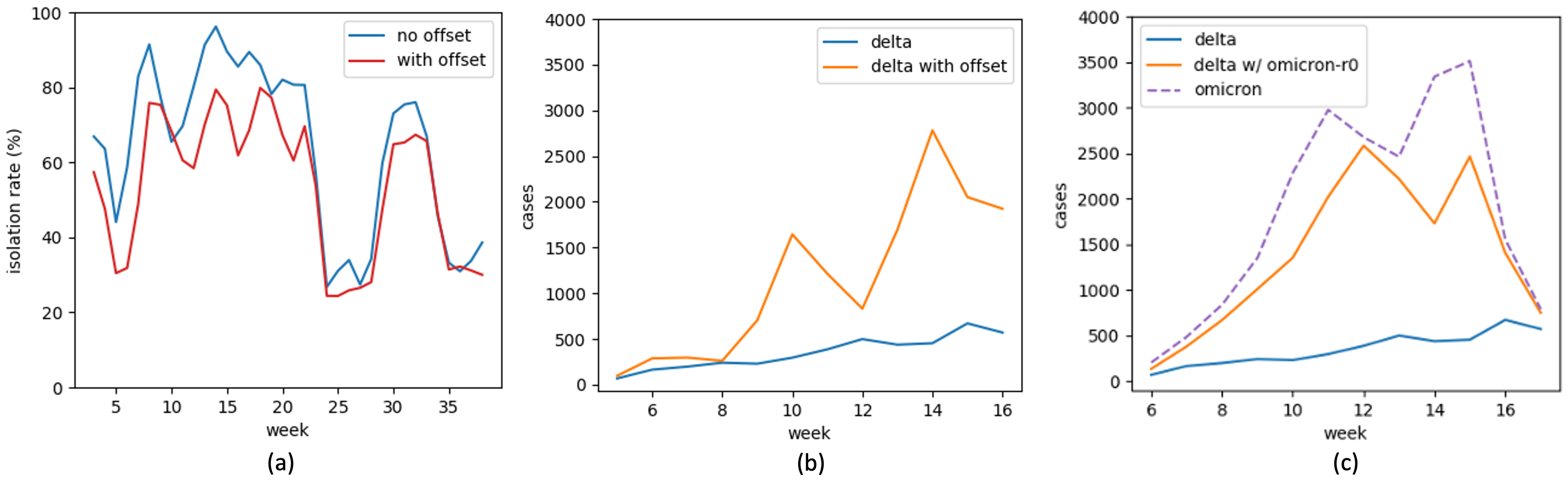}
    \caption{LLM archetypes help explore the interplay between behavior adaptation and environment dynamics in shaping epidemic outcomes. (left) Introducing pandemic fatigue ("the offset") to the prompt reduces relative rates of isolation behavior in the population. (middle) This decrease in isolation behavior translates to increased disease transmission in the population. (right) Comparing the original delta wave (in \textcolor{blue}{blue}), delta wave with "omicron-like" transmissibility (in \textcolor{orange}{orange}) and the omicron wave (shown in \textcolor{mypurple}{dashed purple} to indicate this emerges at a later time). The omicron wave achieves a higher peak than both the original and "omicron-like" delta wave due to coupled impact of viral transmission and time-induced pandemic fatigue.}
    \label{fig:q-counterfactual_sim}
\end{figure*}

\vspace{-4mm}

\section{Counterfactual SIMULATIONS}
\label{sec:bench_analysis}
The Lucas Critique~\cite{lucas1976econometric} posits that historical data alone can never predict what happens when a new policy is adopted, since behavior may adapt while outcomes are realized. By balancing individual agency and simulation scale, LLM archetypes allow us to analyze the relative impact of behavior adaptation and environment dynamics in shaping real world outcomes. This section explores the interplay of pandemic fatigue and variant transmissibility on the severity of disease outbreaks. 

We consider transmission of two variants of COVID-19 - Delta ($\beta=[2.5-4.0]$, April 2021) and Omicron ($\beta=[5.5-8.0]$, November 2021) - which emerged at different stages of the pandemic. While Omicron was roughly 2-3 times more transmissible than Delta, it produced 5-20 times the case intensity~\cite{alam2023beta}. We hypothesize this is due to the coupled dynamics of time-induced fatigue and increased transmissibility. 

Using our model calibrated to the Delta wave, we conduct two counterfactual simulations:

\textbf{Q1: What if we had the delta wave later?}: To simulate time-induced behavior change ("pandemic fatigue"), we update the user prompt with an artificial offset: \textit{"it has been number of \{weeks + OFFSET \} since start of the epidemic"}, where OFFSET is 30 weeks.

\textbf{Q2: What if we had the omicron wave earlier?}: We update $\beta$ in the simulation trace to mimic the Omicron variant while keeping the same behavioral dynamics.

Figure ~\ref{fig:q-counterfactual_sim} illustrates the results of our counterfactual analysis, helping decouple the impact of behavior change and viral transmissibility on disease waves during the pandemic. First, we observe that time-induced fatigue can alter behavior of individuals with agents demonstrating lower willingness to isolate when prompted using the OFFSET(~\autoref{fig:q-counterfactual_sim}(a)). This behavior change can result in a more severe outbreak even with viral dynamics do not change (~\autoref{fig:q-counterfactual_sim}(b)). Second, an early onset of "omicron-like variant" would have been more destructive due to higher transmissibility, but not as severe as the actual omicron wave. The actual Omicron wave was exacerbated by the coupled influence of behavior change (additional "pandemic fatigue" due to extended duration). Such analyses can inform policy decisions during a pandemic helping appropriately allocate resources to both clinical and behavioral interventions. 

We note that archetypes enable such analysis since they provide the ability to query adaptations in individual behavior, via expressive natural language, and also measure the cascading impact of individual decisions at a population scale. These analyses are challenging with other agent architectures due to trade-offs between simulation scale (compromised when using LLMs for each agent) and individual agency (lost when using heuristic agents). We note that for small populations with high-personalized interventions, LLMs-as-agent can be viable architecture (as explored in works like ~\cite{parkGenerativeAgentsInteractive2023, vezhnevetsGenerativeAgentbasedModeling2023a}). Archetypes are useful when analyzing outcomes in large populations with demographic-resolved interventions as often required for policy making.

\section{Discussion}
\label{sec:discussion}
This section addresses implementation considerations and limitations. We analyze sensitivity of Archetype-specific design choice when querying LLMs and introduces an API to generalize utility of our contributions to diverse scenarios. We also discuss some promising future capabilities and summarize limitations of the work.

\begin{figure}[h!]
    \centering
    \includegraphics[width=0.9\linewidth]{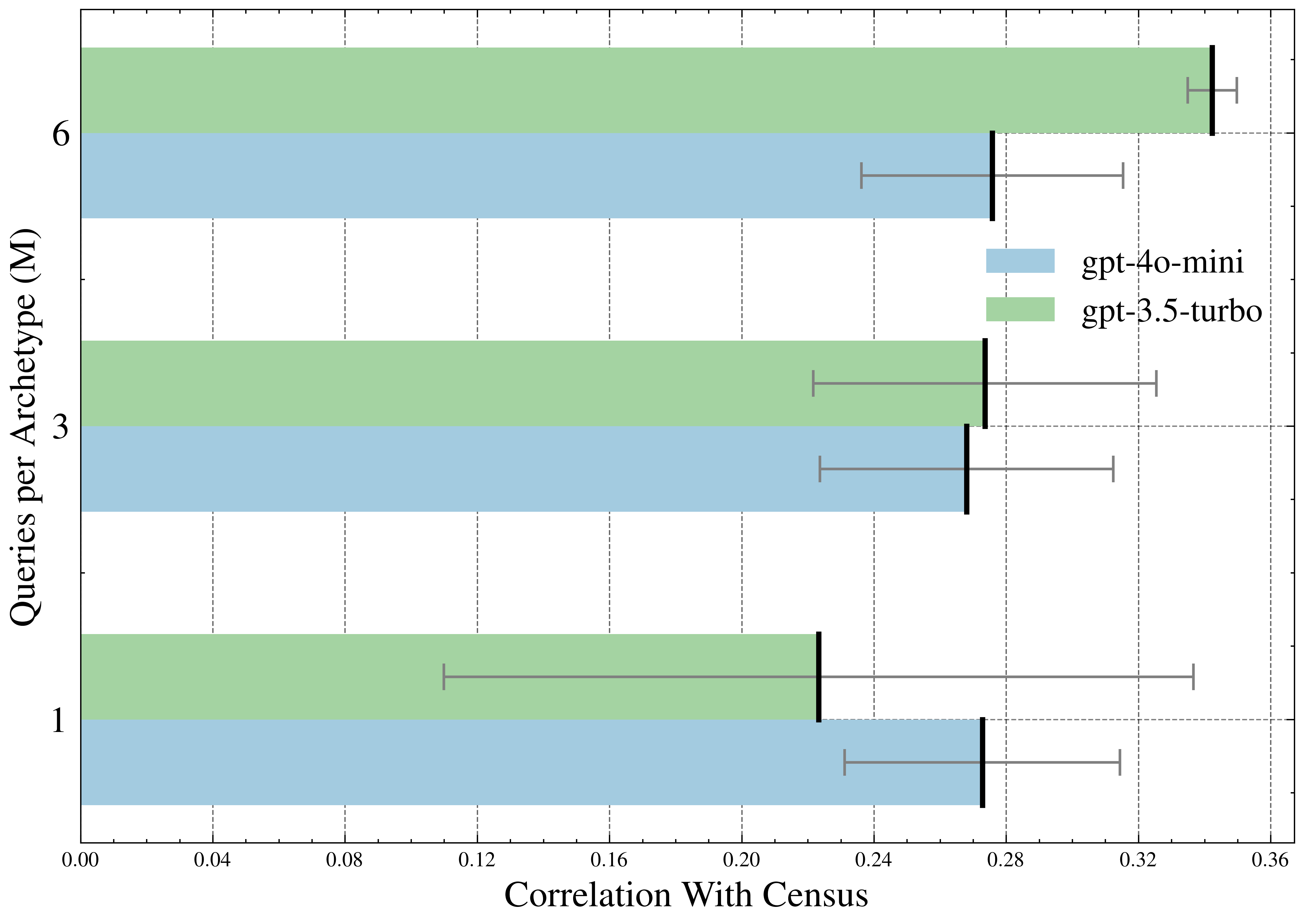}
    \caption{Sensitivity analysis to LLM model choice (GPT-3.5, GPT-4o) and number of queries per archetype (M=1, 3, 6)}
    \label{fig:design-choices}
\end{figure}

\textbf{LLM Consistency}: LLM archetypes are sensitive to the quality and consistency of agent behaviors, which can vary with the choice of choice of model and number of queries per archetype ($M$ in equation~\ref{eq:p_arch}). We repeat the experiment in section~\ref{sec:archetypes} (using Prompt 3) and analyze sensitivity of individual decisions to model choice (GPT-3.5 vs GPT-4o) and number of instances per archetype ($M$=1, 3, 6). Results are shown in ~\autoref{fig:design-choices}. First, for lower $M$, using the superior GPT-4o model improves performance. Our algorithm is agnostic to the choice of LLMs and we anticipate the our results will become progressively better as LLMs mature, making ABMs more reliable. Second, for larger $M$ - when archetype distributions are aggregated over multiple queries - performance significantly improves for GPT-3.5. We hypothesize that this mitigates biased LLM responses and helps capture realistic variability in individual behaviors. For future work, we plan to extensively benchmark different models (both open and closed-source) and design formal guidelines to specify archetype prompts. Finally, we also note that the choice of LLM is not the only factor affecting effectiveness. As demonstrated in Section~\ref{sec:calibration}, population scale is also critical for ABMs where heuristic agents can outperform LLM agents. This highlights the complex interplay between agent sophistication and simulation scale in determining overall ABM utility.


\textbf{Archetype API}: We extend the AgentTorch framework~\cite{chopra2024framework} to integrate an 'Archetype' API, allowing the use of LLMs to prompt agent behavior in large-scale simulations. This API supports both offline and online LLMs, facilitating wider adoption of LLM archetypes in ABMs. We present a code snippet in ~\autoref{fig:archetype-api} and provide additional details including the source code and tutorials in Appendix D. We will also engage with the AgentTorch developers to integrate our API within the core framework.

\begin{figure}[h!]
    \centering
    \includegraphics[width=0.9\linewidth]{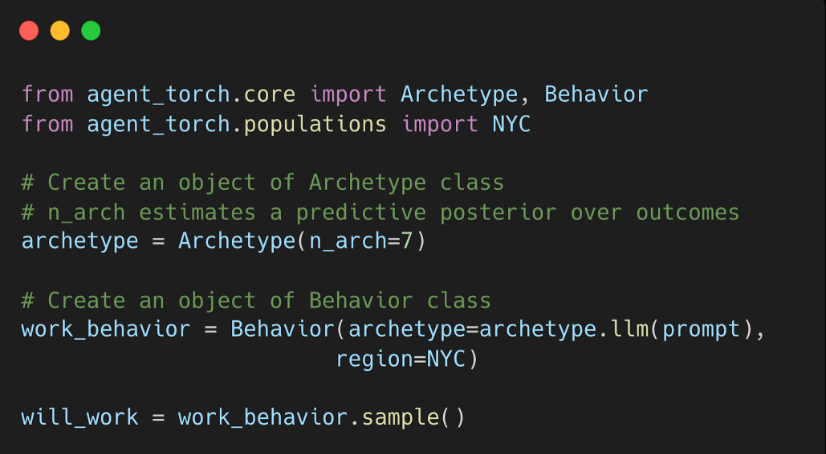}
    \caption{We extend the AgentTorch framework to generalize use of LLM Archetypes. More details in Appendix D.}
    \label{fig:archetype-api}
\end{figure}

\begin{figure}
    \centering
    \includegraphics[width=0.8\linewidth]{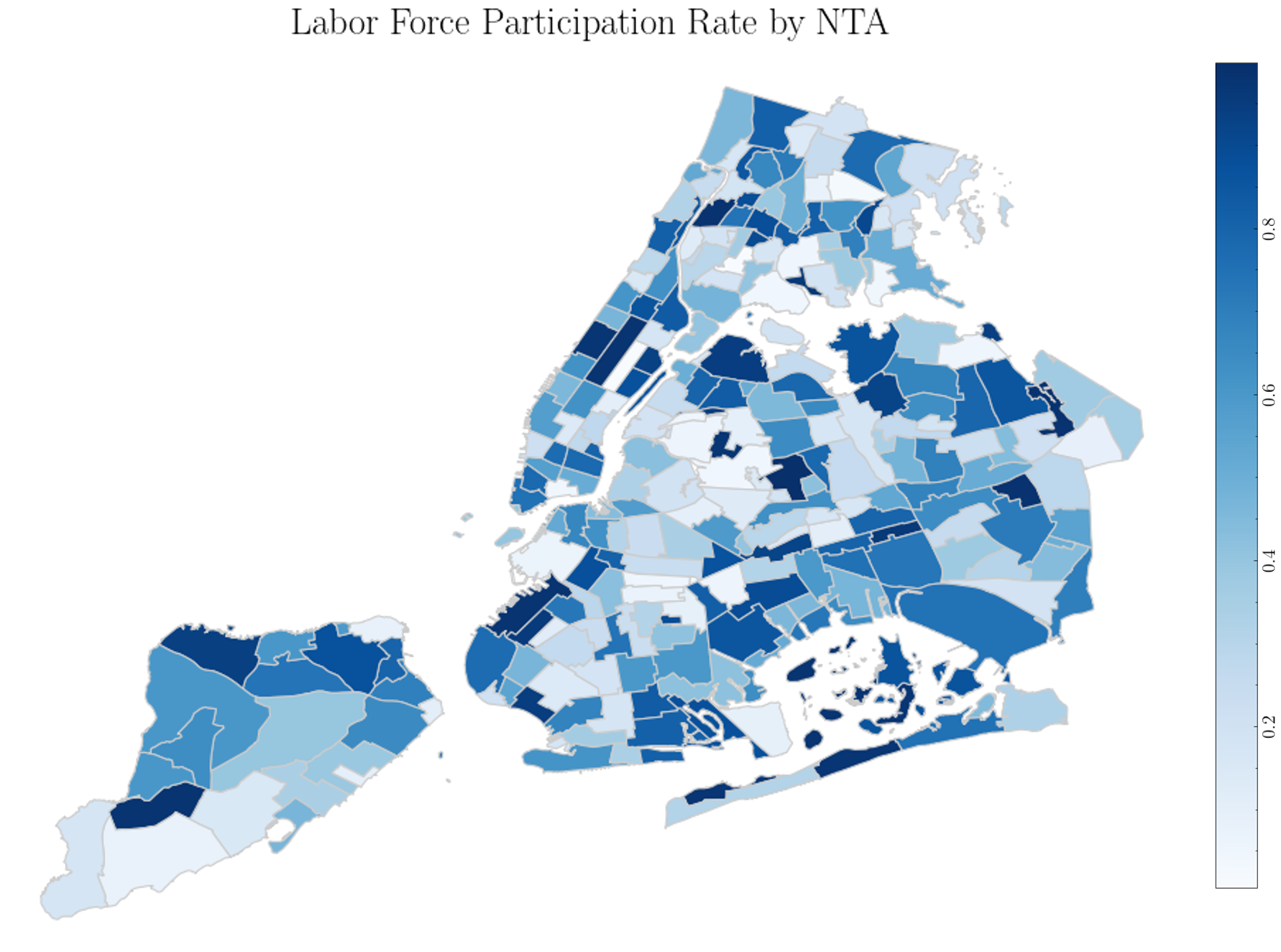}
    \caption{Zip-code level employment behavior for 8.4 million agents in NYC estimated using a model calibrated to coarse borough-level data. LLM Archetypes help overcome limitations of historical data for policy design.}
    \label{fig:historical-limits}
\end{figure}

\textbf{High-resolution Analysis}: LLM archetypes enable measuring granular individual behavior and integrating it with population-scale simulations. Using our model calibrated to borough-level data, we measure the impact of stimulus payments on employment behavior at a granular zip-code level. In future, such analyses can help overcome limitations of historical data in policy design (~\autoref{fig:historical-limits}).

\textbf{Limitations}: While our work demonstrates the potential of LLM archetypes in large-scale ABMs, several challenges and areas for improvement remain. First, ensuring the robustness and fairness of LLM-driven agents remains an open challenge, as LLMs can produce inconsistent or biased outputs, potentially leading to unrealistic agent behaviors. Future work should focus on developing methods to detect and mitigate these biases. Second, while LLM archetypes aid scalability, they may not always capture the desired heterogeneity of individual agents, necessitating the development of more sophisticated archetype selection and interpolation methods. Third, the current implementation focuses on relatively simple agent actions, limiting the complexity of decision-making processes that can be modeled. Extending the action space and implementing multi-scale archetypes could address this limitation. Fourth, the potential for data contamination in LLMs, where models may contain anachronistic information relative to the simulation timeframe, requires careful consideration and the development of techniques to "time-bound" LLM knowledge. Fifth, verifying the accuracy of individual agent behaviors generated by LLMs remains challenging, calling for the development of formal verification methods and benchmarks. We currently measure performance via comparisons with mesoscopic census data and generalization of macroscopic ABM predictions. Despite these limitations, we believe our work represents a significant step forward in agent-based modeling, opening up new possibilities for understanding and addressing societal challenges.

\section{Conclusion}
This work introduces LLM archetypes as a novel approach to scale adaptive agent behavior in large-scale agent-based models (ABMs). By efficiently integrating LLMs into ABMs, we enable the simulation of millions of agents with nuanced, context-aware behaviors while maintaining computational feasibility. Our case study on the COVID-19 pandemic in New York City demonstrates the power of this approach in capturing complex societal dynamics, balancing individual agency with population-scale outcomes. The framework's ability to perform counterfactual analyses addresses key limitations in policy design, offering a powerful tool for tackling real-world challenges. While challenges remain in ensuring robustness and fairness of LLM-driven agents, this work represents a significant step forward in ABM capabilities. By bridging the gap between expressive individual agents and large-scale simulations, our approach opens new avenues for modeling complex systems and informing data-driven policy decisions across various domains.



\begin{acks}
This research was funded by the MIT Media Lab consortium. Arnau Quera-bofarull acknowledges support by the UKRI AI World Leading Researcher Fellowship awarded to Wooldridge (grant EP/W002949/1). M. Wooldridge and A. Calinescu acknowledge funding from Trustworthy AI - Integrating Learning, Optimisation and Reasoning (TAILOR), a project funded by European Union Horizon2020 research and innovation program under Grant Agreement 952215.
\end{acks}



\bibliographystyle{ACM-Reference-Format} 
\bibliography{sample}


\end{document}